\documentclass[12pt]{article}
\usepackage[latin9]{inputenc}
\usepackage{amsmath}
\usepackage{amssymb}
\usepackage{esint}
\usepackage[unicode=true,pdfusetitle,
 bookmarks=true,bookmarksnumbered=false,bookmarksopen=false,
 breaklinks=false,pdfborder={0 0 1},backref=false,colorlinks=false]
 {hyperref}

\makeatletter

\usepackage{slashed}

\usepackage{epsfig}\usepackage{calrsfs}\usepackage{bbm}\usepackage{bm}
\usepackage{color}\usepackage{cancel}\usepackage{mathrsfs}
\usepackage{dsfont}
\usepackage{exscale}

\def\a0size{6}

\newcommand{\lsi}{\raise0.3ex\hbox{$<$\kern-0.75em\raise-1.1ex\hbox{$\sim$}}}
\newcommand{\gsi}{\raise0.3ex\hbox{$>$\kern-0.75em\raise-1.1ex\hbox{$\sim$}}}

\renewcommand{\vec}[1]{{\bf #1}}

\allowdisplaybreaks[1]

\begin{document}
\setlength{\baselineskip}{0.6cm} \global\long\def\figysize{16.0cm}
 \global\long\def\figtopspace{\vspace*{-1.5cm}}
 \global\long\def\figbottomspace{\vspace*{-5.0cm}}

\global\long\def\theequation{\thesection.\arabic{equation}}
\newcounter{saveeqn}
\newcommand{\alphaeqn}{\refstepcounter{equation}\setcounter{saveeqn}{\value{equation}}%
\setcounter{equation}{0}%
\renewcommand{\theequation}{%
        \mbox{\thesection.\arabic{saveeqn}\alph{equation}}}}%
\newcommand{\reseteqn}{\setcounter{equation}{\value{saveeqn}}%
\renewcommand{\theequation}{\thesection.\arabic{equation}}}

\makeatletter \@addtoreset{equation}{section} \makeatother
\renewcommand{\theequation}{\arabic{section}.\arabic{equation}}

\begin{titlepage}

\begin{flushright}
BI-TP 2015/17 
\\
  October 2015 
\\
 
\par\end{flushright}

\begin{center}
\vfill{}

\par\end{center}

\begin{center}
\textbf{\Large{}{{Equilibration, particle production, and self-energy
 }}}{\Large{}{ }} 
\par\end{center}

\begin{center}
\vspace*{0.6cm}

\par\end{center}

\begin{center}
D.~B\"odeker
\footnote{bodeker@physik.uni-bielefeld.de%
}, M.~Sangel%
\footnote{msangel@physik.uni-bielefeld.de%
}, and M.~W\"ormann
\footnote{mwoermann@physik.uni-bielefeld.de%
} 
\par\end{center}

\begin{center}
{\em Fakult\"at f\"ur Physik, Universit\"at Bielefeld, 33501 Bielefeld,
Germany } 
\par\end{center}

\begin{center}
\vspace*{1cm}

\par\end{center}

\begin{center}
\textbf{Abstract} 
\par\end{center}

\vspace{0.5cm}

The discontinuity, or imaginary part of a self-energy at finite
temperature is proportional to the rate at which the corresponding
particles are produced when very few of them are present, and also to
the rate at which their phase space density approaches the
thermal one.  These relations were suggested by Weldon
\cite{weldon-simple}, who demonstrated them for low orders in
perturbation theory. Here we show that they are valid at leading order
in a linear coupling of the produced particles, and to all orders in all
other interactions of the hot plasma, if there is a separation of the
time scales for the production and for the thermalization of the bulk
of the plasma.

\noindent \vspace{0.5cm}

\noindent 
\vspace{0.3cm}

\noindent \vfill{}
 \vfill{}

\noindent \end{titlepage}


\section{Introduction}\label{sec:intro}

The phase space density, or occupancy $f _ \vec{p} (t,\vec{x} ) $
plays a central role in statistical mechanics, both for equilibrium
and non-equilibrium.\footnote{We consider a finite
  volume $ V $, so that the momenta $ \vec p $ are discrete.
  As usual $ V ^{ -1 } \sum _ { \vec p } \to ( 2 \pi ) ^{ -3 } \int d
  ^ 3 p $ in the infinite volume limit.} It is a sensible concept only
for particles which are sufficiently weakly interacting. It is of
little use, e.g., for quarks when the temperature is near or below the
QCD-temperature $T\sim$ 160 MeV. On the other hand, it can make
perfect sense for neutrinos at the same temperature.

In thermal equilibrium the phase space density is given by the
Bose-Einstein or Fermi-Dirac distribution. For non-equilibrium $f$ has
been considered in various situations. Two of them are of interest to
us here: $(i)$ $f$ is close to
its equilibrium value, $|f-f^{{\rm eq}}|\ll1$, 
and $(ii)$  $f$ is very small, $f\ll1$. We restrict ourselves
to homogeneous systems, i.e., $ \vec x $-indepedent $ f $.

We consider the occupancy $ f $ of a particle species $ \Phi $, and we assume
that most physical quantities equilibrate much faster than $ \Phi $. This is
the case if the coupling of $ \Phi $ is much weaker than the couplings of the
interactions of the other degrees of freedom. There may be other quantities $
X _ a $ besides $ f $ which equilibrate slowly as well.  We choose them such
that $ X _ a = 0 $ in equilibrium.  

In equilibrium the slow variables
fluctuate around their thermal expectation values. We will be concerned with
non-equilibrium states for which $ f $ is much larger than a typical thermal
fluctuation.  Then we may consider systems out of equilibrium for which the
out of equilibrium state is specified by the values of $ f $ and the $ X _ a
$, and of the temperature $ T $.\footnote{For simplicity we assume that there
  are no chemical potentials present.}  This description is valid on time
scales much larger than the equilibration time of the fast degrees of
freedom.  The time derivatives of $ f $ and the $ X _ a $ are fully determined
by $ f $, the $ X _ a $, and $ T $, because the state of the system is fully
specified by these quantities.

First consider a system in which both  $f$ and the $ X _ a $ are
close to equilibrium.
The interaction with the thermalized plasma will bring $f$ closer
to equilibrium. Thus for sufficiently small $\delta f \equiv f - f ^ {\rm eq }$
and $X_a$ we must have a linear relation 
\begin{align}
   \dot f _ {  \vec p , \lambda    } 
   =-
   \sum _ { \vec p ',\lambda ' } 
   \widetilde{ \Gamma } _ {\vec{p}\vec{p}', \lambda  \lambda  ' }
   \delta f _  {\vec{p}', \lambda  '} 
   + \cdots 
   \label{equieq}
   ,
\end{align}
where $ \lambda  $ labels possible spins or helicities. Furthermore, 
`$ + \cdots $' denotes terms  linear in  the other slow variables, as well has 
higher powers of the deviations from equilibrium. 
We will see that at leading order in the interaction of $\Phi$
\begin{align}
  \widetilde{ \Gamma } _ { \vec{p}\vec{p}', \lambda  \lambda  ' } 
  =
  \delta  _ { \vec p \vec p' } \delta  _ { \lambda  \lambda  ' } 
  \Gamma_{\vec p , \,\lambda}^{\rm eq}
   \label{Gamdel}
   .
\end{align} 
Therefore  we can  write
\begin{align}
   \dot f _ { \vec p, \lambda  }
   = - \Gamma^{\rm eq} _ { \vec p, \lambda  }
   \, \delta f  _ { \vec p, \lambda  }
   + \cdots 
   \label{eff} 
   .
\end{align}

Now  consider an out-of-equilibrium system with 
 very few $ \Phi  $-particles present, {\it  i.e.} 
$f\ll1$. 
The interactions of the otherwise thermalized plasma will then
create $ \Phi $-particles. At the same time the interactions can drive the rest 
of the plasma slightly away from equilibrium.  Thus, in general one would have to
consider the evolution not only of $f$ but also of the $ X _ a $.
However, as long as these are still sufficiently small,
we may  expand around $f=0$ and $X_a=0$, and keep only the lowest order term
in this expansion, which  defines the production rate $ \Gamma ^  {\rm pro} $
of~$\Phi  $, 
\begin{align}
   \dot f 
   =\Gamma ^  {\rm pro}
    +
    \cdots
   \label{production} 
   .
\end{align}
Now  `$ + \cdots $' denotes terms of linear or higher order in $ f $
and $ X _ a $.  The production rate 
depends on $|\vec{p}|$ 
and the parameter which characterizes the equilibrated plasma, i.e., the
temperature.

If (\ref{eff}) were also valid  for very small $ f $,  where 
$ \delta  f $ is {\it not} small, but instead of order one, 
$ \delta  f \simeq - f ^ {\rm eq } $, we would have
\begin{align}
   \Gamma  ^ {\rm pro } = f^{\rm eq } \Gamma  ^  {\rm eq }
   \label{proeq} 
   .
\end{align} 
Weldon \cite{weldon-simple} has shown that this relation holds at
leading order in the couplings, and also for multiparticle processes, assuming the
validity of the Boltzmann equation. With the same assumptions he found 
that the two rates are proportional to the discontinuity of the $ \Phi 
$-self-energy, see (\ref{gammaself}).
In~\cite{nrlg} it was
assumed that (\ref{proeq}) also holds when radiative corrections 
\cite{laine-nr,salvio} are
taken into account. However, for a proper treatment of radiative
corrections one should justify (\ref{proeq}) 
beyond leading order, where the Boltzmann equation may no longer be valid.

We write the Hamiltonian as 
\begin{align} 
  H = H _ 0 + H _ {\rm int } 
  \label{H} 
  ,
\end{align} 
where $ H _ 0 $ describes free $ \Phi $-fields and all other fields including
their interactions. The interaction of $ \Phi $ with the other fields is
contained in $ H _ {\rm int } $. The relation between $ \Gamma ^ {\rm pro } $
and the discontinuity of the self-energy is well
known to be valid at leading order in $ H _ {\rm int } $ and to {\it all} orders in
all other interactions contained in $ H _ 0 $ \cite{weldon-dilepton,asaka}.
Recently a non-linear evolution equation for the phase space density of
sterile neutrino dark matter particles which does not use a perturbative
expansion in $ H _ 0 $ has been obtained using an ansatz for the
non-equilibrium density matrix \cite{ghiglieri-improved}. The relation
(\ref{proeq}) can then be obtained by expanding the  equation of
\cite{ghiglieri-improved}
around small $ f $.  

In this note we show that (\ref{proeq}) holds to all orders in the couplings
in $ H _ 0 $ and to leading order in $ H _ {\rm int } $, if there is a
separation of the time scales associated with the interactions in $ H _ 0 $
and those in $ H _ {\rm int } $. We use the theory of quasi-stationary fluctuations
\cite{landau5} without making any ansatz for the density matrix, and we treat
both bosons and fermions.  Our assumptions differ somewhat from
\cite{weldon-simple}, where the production of $ W $-bosons was considered;
for this process our assumption about the separation of time scales is not
satisfied.

\section{Charged particle species}\label{s:dirac} 
Equilibration rates of slowly evolving quantities
can be computed from quantum field theory by 
defining appropriate number density operators and matching their real 
time correlation functions with the corresponding one computed from the
effective kinetic equations like (\ref{equieq}) \cite{landau5,bodeker-washout}.

First we need an operator expression for the occupancy.  We will treat
$ H _ {\rm int } $ as a small perturbation and work at leading order
in $ H _ {\rm int } $.  In the interaction picture with respect to $ H
_ {\rm int } $ the field operator $ \Phi $ can be written as
\begin{equation}
   \Phi_{\rm I}(x)
   = 
    \sum_ { \vec p, \lambda  } 
    \frac 1 { \sqrt{ { 2  E _ \vec p } V} }
   \left [e ^{ -ipx } \,   u_{ \vec p, \lambda }    c_ { \vec p, \lambda } 
     +
      e ^{ ipx } \, v_{ \vec p, \lambda }  d ^\dagger _ { \vec p, \lambda } 
     \right ] _ { p ^ 0 = E _ \vec p } 
     \label{phiI}
     ,
\end{equation}
which defines
annihilation operators $ c _ {\vec{p},\lambda} $, and $
d_{\vec{p},\lambda} $ for particles and antiparticles, 
normalized such that
\begin{equation}
 [c_{\vec p ,\lambda},c_{\vec p' ,\lambda'}^\dagger]_{-\sigma}
 =
  \delta  _ { \vec p \vec p' } \delta_{\lambda\lambda'}
 \label{aadcomm}
 , 
\end{equation}
and similarly for the antiparticles. Here $\sigma=+1$ for bosons and
$\sigma=-1$ for fermions.
The occupancy operator for particles in the interaction picture can be
defined as
\begin{align}
  \left (   f_{\vec{p,}\lambda } \right ) _ {\rm I } 
  & 
  \equiv
  c_{\vec{p},\lambda}^{\dagger}c_{\vec{p},\lambda}
  \label{fint} 
  . 
\end{align}
We will be interested in the Heisenberg picture-$ f $ which is
related to (\ref{fint}) via $ f = e ^{ i H t } e ^{ -i H _ 0 t} f _
{\rm I } \,e ^{ i H _ 0 t } e ^{ -i H t } $.  For non-interacting $
\Phi $ the operator $ f_{\vec{p,}\lambda } $ would be conserved and
would precisely be the occupancy of free particles.  It is, however,
also defined for non-vanishing $ H _ {\rm int } $.

Following~\cite{bodeker-washout} one can then compute
the coefficients in (\ref{equieq}) 
via 
\begin{equation}
   \widetilde{ \Gamma } _{\vec{p}\vec{p}', \lambda  \lambda  ' } 
   =
   \frac{1}{2 V} 
   \sum _ { \vec p '', \lambda  ''  } 
   \lim _ { \omega  \to 0 } \frac { 
     \rho_ {\vec p \vec p '', \lambda  \lambda  '' }(\omega)
     } \omega  
   \left(\Xi^{-1}\right)_{\vec p '' \vec p ' , \lambda  '' \lambda  '} 
  \label{matching}
\end{equation}
with the spectral function 
\begin{equation}
     \rho_ {\vec{p} \vec p' , \lambda  \lambda  ' } (\omega)
       = 
       \int dt\; e^{i\omega t} \left \langle 
         \left[\dot{f}_{\vec{p}, \lambda  }(t),
           \dot{f}_{\vec{p}', \lambda  '}(0)\right]
       \right\rangle 
       \label{spec}
       ,
\end{equation}
and the matrix
\begin{equation}
   \Xi_ { \vec{p} \vec{p}', \lambda  \lambda  ' }
   \equiv 
   \frac{1}{TV}
   \left \langle\delta f_{\vec{p}, \lambda  }\,\delta f_{\vec{p}', \lambda'  
   }\right \rangle
   \label{sus}
   .
\end{equation}
At leading order in $ H _ {\rm int } $ the right-hand side of
(\ref{sus}) is determined by the free theory which yields  
\begin{align} 
  \left \langle\delta f_{\vec{p}, \lambda  }\,\delta f_{\vec{p}', \lambda'  
   }\right \rangle
   = 
   \delta_{\lambda\lambda'}
   \delta_{\vec p\vec p'}
   f_{{\rm \sigma}}(E_{\vec{p}}
)
   \Big [ 1+\sigma 
   f_{{\rm \sigma}}(E_{\vec{p}}
) \Big ]
 \label{freesus}
 .
\end{align} 
Here $f_{{\rm \sigma}}$ is the Bose-Einstein- or Fermi-Dirac-distribution
for $\sigma=+1$ or~$\sigma=-1$ respectively, and $ E
_ \vec p  = ( \vec p ^ 2 + m _ \Phi  ^ 2 ) ^{ 1/2 } $. 

To compute the time derivatives in (\ref{spec}) we need to specify the interaction
of $ \Phi  $. 
We consider a linear coupling 
\begin{equation}
    {\cal L}  _ {\rm int } 
    =
    - \overline{ J }  \Phi 
    -
               \overline{ \Phi  }  J
    \label{Hint}
    ,
\end{equation}
where the operator $J$ can be elementary or composite and does not
contain $ \Phi $. Furthermore, $ \overline \Phi \equiv \Phi ^\dagger $
for integer spin, and $ \overline \Phi \equiv \Phi ^\dagger \gamma ^
0$ for spin 1/2.

In the following we will obtain a master-formula (\ref{gamma}) for the
coefficients $\widetilde{ \Gamma } _{\vec{p}\vec{p}', \lambda \lambda
  ' }$ 
at leading order in $ H _ {\rm int } $.  For this purpose we now
determine the spectral function \eqref{spec} to this order.
The time derivatives  of the occupancy operators in (\ref{spec}) are 
given by
\begin{equation}
  \dot{f}_{\vec{p},\lambda}
  =i[H,{f}_{\vec{p},\lambda}]
  \label{werner}
  .
\end{equation}
To simplify (\ref{werner}) we rewrite the right-hand side in terms of
interaction picture operators, use (\ref{phiI}) and (\ref{aadcomm}) and finally
express the result in terms of Heisenberg operators. At leading order in $ H _
{\rm int } $ this yields 
\begin{align}
  \dot{f}_{\vec{p},\lambda}
  =  \frac i { \sqrt{ 2  E _ p V} } \int d^{3}x\left[\,
    \overline J(x)e^{-ip x }u_{\vec{p},\lambda} c_{\vec{p},\lambda} 
    - \mbox{   h.c.}\right]
  \label{fdot} 
  .
\end{align}
Since we work at leading order in $ H _ {\rm int } $, we neglect it in
the computation of the expectation value in (\ref{spec}).  Now we
insert (\ref{fdot}) into (\ref{spec}), use (\ref{aadcomm}) and treat $
\Phi $ as a free field, after which  we obtain  Wightman functions 
\begin{align} 
    \Delta_{AB}  ^ > ( x ) \equiv  \langle A  ( x ) B ( 0 ) \rangle, 
    \qquad
    \Delta_{AB}  ^ < ( x ) \equiv  \sigma  \langle B ( 0 ) A  ( x ) \rangle 
    \label{wightman} 
    , 
\end{align} 
of $ A = \overline u _ { \vec p, \lambda  } J $ and $ B = A ^\dagger   $.
Using
\begin{align} 
  \Delta_{AB}  ^ > ( \omega  ) = \Big [ 1 + \sigma  f _ \sigma  ( \omega  ) 
  \Big ]  \widetilde{  \rho  }_{AB}  ( \omega  ) 
  , \qquad 
  \Delta_{AB}  ^ <  ( \omega  ) =  \sigma  f _ \sigma  ( \omega  ) 
   \widetilde{ \rho }_{AB}   ( \omega  ) 
   \label{ws} 
\end{align} 
with the spectral function 
\begin{equation}
   \widetilde{ \rho  }_{AB} (p ) 
   =
   \int d^{4}x\,e^{ipx}
   \left\langle \left[A  (x),B (0 ) \right]_{-\sigma}\right\rangle 
   \label{rhotilde} 
\end{equation}
we obtain after some work 
\begin{equation}
      \lim _ { \omega  \to 0 } \frac { 
     \rho_ {\vec p \vec p ', \lambda  \lambda  ' }(\omega)
     } \omega  
     =
     \frac{ 1
     }   { T E _ \vec p }
       \delta_{\vec{p} \vec{p}'}    \delta_{\lambda\lambda'} 
     f_{\sigma}(E_{\vec{p}}
  )
  \Big [  1+\sigma 
    f_{\sigma}(E_{\vec{p}}
    ) \Big ]
    \, 
    \widetilde{\rho}_{\overline u  
      J ,  \overline J u
    }(E_{
    \vec{p}},\vec{p}) 
    \label{work} 
    .
\end{equation}
Using \eqref{matching} in combination with (\ref{freesus}) and
(\ref{work}), we find that the equilibration rate indeed takes the form
(\ref{Gamdel}) with
\begin{equation}
  \Gamma  ^  {\rm eq }  _{\vec{p}, \lambda   }  
  =
  \frac { 1} { 2 E _ \vec p } 
  \,
  \, \widetilde{\rho}_{\overline u  
    J  , \overline J u
  }
      (E_{\vec{p}},\vec{p} )
   .
   \label{gamma}
\end{equation}

Like in \cite{weldon-simple} 
we can relate this expression to the $ \Phi  $-self-energy. 
A spectral function such as (\ref{rhotilde}) can be obtained
by starting with an imaginary time,  or Euclidean correlator
\begin{equation}
   \Delta  _{AB}(i \omega  _{n},\vec{p})
   \equiv 
   \int_{0}^{\beta} d\tau\int d^{3}x
   \, e^{i ( \omega  _{n}\tau  -\vec{p}\cdot\vec{x} ) }
   \left\langle A (-i \tau , \vec x) B(0)\right\rangle 
   \label{euclid} 
   ,
\end{equation}
with 
the discrete Matsu\-bara frequencies $ \omega _ n = n \pi T $ with even
and  odd integer $ n $ for bosons and fermions, respectively. 
(\ref{euclid}) 
can be continued   $ i \omega  _ n \to p ^ 0 $ into the complex
$ p ^ 0 $-plane. The
resulting function has cuts and poles on the real $ p ^ 0 $-axis. The 
spectral function  is then proportional to its 
discontinuity $ \text{Disc}
  \Delta  ^{{\rm }}(p ^ 0) \equiv 
  \Delta  ^{{\rm }}(p ^ 0+i0 ^ +  ) 
  - \Delta  ^{{\rm }}(p ^ 0- i 0 ^ + ) 
$ across the real axis, 
\begin{equation}
  \widetilde{\rho}_{A B}(p ^ 0  ,\vec{p})
  =
  \frac{1}{i}\, \text{Disc}
  \Delta  _{AB}^{{\rm }}(p ^ 0  ,\vec{p})
  .
\end{equation}
At leading order in $ H _ {\rm int } $ 
the Euclidean $\Phi$-self-energy 
is given by
\begin{equation}
 \Sigma   (i \omega  _ n,\vec{p})
 =
 \int_{0}^{\beta}d\tau\int d^{3}x
 \, e^{i (  \omega  _ {n}\tau  - \vec{p}\cdot\vec{x}) }
 \left\langle J(-i \tau  , \vec x) \,\overline{ J } (0)\right\rangle
 \label{self} 
 .
\end{equation}
Therefore
\begin{align}
   \Gamma_{\vec{p},\lambda} ^{ \rm eq } 
   = & \frac{1}{2iE _ \vec p} 
   \overline u _{\vec{p},\lambda}
   {\rm Disc  } \, 
   \Sigma ^{ \rm  } (E _ \vec p , \vec{p})u_{\vec{p},\lambda}
   \label{gammaself}
   .
\end{align}
The same relation has been obtained by Weldon \cite{weldon-simple}:
the equilibration rate of particles is proportional to the
discontinuity of their self-energy. We differ from Weldon in two
respects: $ ( i ) $ we have shown that this result is valid at leading
order in $ H _ {\rm int } $ and to {\it all} orders in the other
interactions. In particular, we have, unlike~\cite{weldon-simple}, not
made use of the Boltzmann equation which is not valid beyond leading
order. $ (ii) $ We have assumed a separation of the time scales on
which $ \Phi $ thermalizes and on the other thermalization times in
the system. Our derivation would not be valid for the equilibration of
electroweak gauge bosons, which was considered by Weldon.

The relation between the production rate (\ref{production}) and the
self-energy or the correlation functions of $ J $, which is valid at
leading order in $ H _ {\rm int } $ and to all orders in $ H _ 0 $, is
well known~\cite{weldon-dilepton,asaka}. It can be obtained by
considering the probability $ P ( t ) $ to find one $ \Phi $-particle
with momentum $ \vec p $ and spin $ \lambda $ at time $ t $ when there
was none at time $ t = 0 $. In the interaction picture it can be written as
\begin{align}
   P ( t ) = \Big | 
      \langle f ; \vec p, \lambda | U _ {\rm I }  ( t , 0) 
       | i  \rangle  
       \Big |^ 2
       \label{prob} 
       ,
\end{align} 
where $ i $ and $ f $ label states which contain no $ \Phi 
$-particles.  Now expand the time evolution operator $ U _ {\rm I }$
in powers of $ H _ {\rm int } $, and choose $ t $ small enough, 
$ t \ll t _ {\rm int } $ with some time scale $ t _ {\rm int } $,  so that
it is sufficient to keep only the term linear in $ H _ {\rm int } $.
Using (\ref{Hint}), summing over $ f $, and thermally averaging  over
$ i $ turns (\ref{prob}) into
\begin{align}
  \langle P ( t ) \rangle 
  = 
  \frac 1 { 2 E _ { \vec p } V }
  \int  _ 0 ^ t d t _ 1 \int _ 0 ^ t d t _ 2 
  \int d ^ 3 x _ 1 \int d ^ 3 x _ 2 \,
  e ^{ i p ( x _ 1 - x _ 2 ) } 
  \left \langle 
    \overline J u ( x _ 2  ) \overline u J ( x _ 1  ) 
    \right \rangle 
    \label{avprob} 
    .
\end{align} 
Using translational invariance of the thermal average and shifting integration
variables this becomes
\begin{align}
   \langle P ( t ) \rangle 
   =  
  \frac 1 { 2 E _ { \vec p }  }
   \int _ 0  ^ t d t _ 2 \int _ { -t _ 2 } ^{ t - t _ 2 } d t _ 1
   \int d ^ 3 x _ 1 \,
   e ^{ i p x _ 1 }
     \left \langle 
    \overline J u (0 ) \overline u J ( x _ 1  ) 
    \right \rangle 
   \label{shift} 
   .
\end{align} 
Now one has to assume that the Wightman function on the
right-hand-side has a finite correlation time $ t _ {\rm corr } $, so
that it practically vanishes for times $ | t _ 1 | > t _ {\rm corr }
$, and that furthermore $ t _ {\rm corr } \ll t _ {\rm int }
$.\footnote{This step is usually omitted.} Then we can choose $ t $ in the range
$ t _ {\rm corr } \ll t \ll t _ {\rm int } $, so that $ t _ 2 $
and $ t - t _ 2 $ are typically much larger than $ t _ {\rm corr } $.
Therefore the $ t _ 1 $-integral hardly changes if the integration
limits are replaced by $ \pm {\infty } $, which gives
\begin{align}
   \langle P ( t ) \rangle
   \simeq   
   t 
   \frac{\sigma}{2 E_{\vec{p}}} 
   \Delta  ^{ < } _ { \overline u  J,\overline J u} ( E _{\vec{p}}, \vec p )
   \label{probl} 
   .
\end{align} 
The production rate is by definition
\begin{align} 
   \Gamma  ^{ \rm pro }_{\vec{p}, \lambda  } 
   = 
   \frac { \langle P ( t ) \rangle } 
   { t V ( 2 \pi  ) ^ { - 3 } d ^ 3 p } 
   \label{gamdef} 
   ,
\end{align} 
which in finite volume equals $ \langle P ( t ) \rangle /t $. 
Therefore 
\begin{align}
   \Gamma  ^{ \rm pro }_{\vec{p}, \lambda  } 
   = 
   \frac{\sigma}{2 E_{\vec{p}}} 
   \Delta  ^{ < } _ { \overline u  J,\overline J u} ( E _{\vec{p}}, \vec p )
   \label{gampro}
   .
\end{align}
Combining this with  (\ref{ws}) and (\ref{gamma}) one sees that the relation
(\ref{proeq}) indeed holds, again to leading order in $ H _ {\rm int } $
and to all orders in $ H _ 0 $.

\section{Uncharged particles} \label{s:majorana}
Now consider particles which are their own antiparticles.  Important
examples could be sterile neutrinos which may be responsible for the
baryon asymmetry of the Universe~\cite{fukugita} or constitute the
dark matter \cite{bulbul}, or other interesting dark matter candidates, or
photons produced in a quark-gluon plasma.  In this case one can write
$ \Phi $ as a field which is equal to its charge conjugate
\begin{align}
  \Phi  = \Phi  ^ C
  ,
\end{align} 
where $ \Phi ^ C \equiv S ( \overline \Phi {} ) ^ T $ with an
appropriate matrix $ S $. Then one can write
\begin{equation}
    {\cal L}  _ {\rm int } 
    =
    - \overline{ I }  \Phi 
    = -
               \overline{ \Phi  }  I
    \label{majoint}
    .
\end{equation}
In this case we find 
\begin{equation}
    \dot f_{\vec{p},\lambda ,\rm uncharged}
    =
    \frac i { \sqrt{ 2  E _ p V} } \int d^{3}x
      \overline I(x)
      \left[\,
    e^{-ip x }u_{\vec{p},\lambda} c_{\vec{p},\lambda} 
    - e^{ip x }v_{\vec{p},\lambda} c_{\vec{p},\lambda} ^\dagger  
  \right]
    \label{fdotmajo} 
    .
\end{equation}
Now we insert this into (\ref{spec}) and follow the same steps as in section
\ref{s:dirac}. We obtain correlation functions of $ \overline u I $
and of $ \overline v I $. Using $ v = u ^ C $ and the fact that
(\ref{majoint}) is hermitian we obtain
\begin{equation}
  \Gamma  ^  {\rm eq }  _{\vec{p}, \lambda   }  
  =
  \frac { 1} { 2 E _ \vec p } 
  \,
  \, \widetilde{\rho}_{\overline u  
    I  , \overline I u
  }
      (E_{\vec{p}},\vec{p} )
   .
   \label{gammamajo}
\end{equation}
The $ \Phi  $-self-energy is
the same as (\ref{self}) but 
with $ J $ replaced by $ I $. Thus we obtain precisely the same 
relation  (\ref{gammaself})
between the discontinuity of the self-energy and the equilibration rate as
we did for charged particles. Also 
the production rate is simply obtained by replacing $ J $ by $ I $
in (\ref{gampro}) which shows that (\ref{proeq}) holds under the same 
conditions as in the charged particle case.

%
In the case of sterile neutrinos which are uncharged particles, 
the interaction is
usually written in the form (\ref{Hint}). In the symmetric phase of
the Standard Model $J=\widetilde{\varphi}^\dagger h_{\nu} \ell$, where
$\widetilde{\varphi}=i\sigma^{2} \varphi^{*}$ is the SU(2) conjugate
Higgs doublet, $\ell$ are the (left-handed) SU(2) lepton doublets and
$h_{\nu}$ is the Yukawa coupling matrix. In the symmetry-broken phase
the dominant contribution is obtained by replacing $ \widetilde{
  \varphi } $ by its expectation value which gives $ J = v h _ \nu \nu 
/\sqrt{ 2 } \label{Jbroken} $ with $ v = 246 $GeV and the
neutrino-field $ \nu $.  Writing $ \Phi   $ as a Majorana spinor, 
$ \Phi   ^ C = \Phi   $, one can also write the interaction in the form
(\ref{majoint}), where now 
$ 
  I=J+ J ^ C
  \label{I} 
  .
$ 
To compute the
equilibration rate one can use either (\ref{Hint}) or (\ref{majoint}).
In the latter case one can directly use the result (\ref{gammamajo})
which shows that (\ref{gammaself}) again holds.

For practical calculations it may be more convenient to write the
rates in terms of correlation functions of $ J $.  To achieve this,
one may either use (\ref{gammamajo}) and re-express $ I $ in terms of
$ J $, or directly start from the form (\ref{Hint}).  Using that
correlators like $\langle JJ\rangle $ in the Standard Model vanish due
to $B-L$ conservation one obtains\footnote{It is interesting to note
  that the second term in the square bracket on the right-hand side
  would give precisely the equilibration rate for anti-fermions in the
  charged particle case (see section \ref{s:dirac}).  }
\begin{equation}
    \Gamma  ^  {\rm eq }  _{\vec{p}, \lambda   }  
  =
  \frac { 1} { 2 E _ \vec p } 
  \Big [ 
  \, \widetilde{\rho}_{\overline u  
    J  , \overline J u
  }
      (E_{\vec{p}},\vec{p} )
   +
   \widetilde{ \rho  }_ { \overline{ v } 
      J ,\overline J \,  v 
    }(-E_{\vec{p}},-\vec{p})
   \Big ]  
   .
   \label{gammac}
\end{equation}
Averaging over spins/helicities, keeping in mind that
$ J $ is left-handed, one obtains 
\begin{equation}
  \Gamma_{\vec{p}}^{\rm eq}
  =
  \frac{1}{4E_{\vec{p}}}\text{Tr}
  \left \{ \slashed
    p\left[\widetilde{\rho}_{J\overline{J}}(E_{\vec{p}},\vec{p})
      +\widetilde{\rho}_{J\overline{J}}(-E_{\vec{p}},-\vec{p})\right]
  \right \} 
  .
  \label{sterilerate}
\end{equation}
Combined with (\ref{proeq}) this gives 
the result for the production rate which was obtained  in \cite{asaka} 
at leading order in $ H _ {\rm int } $ and to all orders in $ H _ 0 $.

\section{Summary and discussion}
\label{s:summary}
We have revisited Weldon's relations~\cite{weldon-simple} between the
equilibration rate of a particle species $ \Phi $,
its thermal production rate, and the 
discontinuity of its self-energy. 
To obtain these relations we had to assume, in contrast
to~\cite{weldon-simple}, that $ \Phi $ equilibrates much more slowly
than the bulk of the plasma. 
We find that these relations are valid to leading
order in the interaction of $ \Phi $ and to {\it all} orders in the other
interactions.  Unlike ~\cite{weldon-simple} we did not make any use of
a Boltzmann equation, which is only valid at leading order.
Our results imply that radiative corrections to the production rate
of sterile neutrinos~\cite{laine-nr,salvio} can also be used for the
equilibration rate, which has been implicitly
assumed previously when incorporating radiative corrections into
leptogenesis computations \cite{nrlg}.

In \cite{ghiglieri-improved} a non-linear evolution equation for the phase
space density of sterile neutrino dark matter particles
has been obtained using an {\it ansatz} for the non-equilibrium density matrix. If
this equation is expanded around  $ f=0 $ one also finds the relation
(\ref{proeq}) between equilibration and production rates. It would be
interesting to see whether also higher order terms in the non-linear evolution
equation for the phase space density of sterile neutrino dark matter particles
of \cite{ghiglieri-improved} can be obtained with the methods we used here.




\appendix
\global\long\def\theequation{\thesection.\arabic{equation}}


\end{document}